\documentclass[aps, prl, reprint, superscriptaddress, floatfix]{revtex4-2}

\usepackage{graphicx} 
\usepackage{braket}
\usepackage{physics}
\usepackage{amsmath}
\usepackage{amssymb}
\usepackage{mathtools}
\usepackage{hyperref}
\usepackage[capitalize]{cleveref}
\usepackage{xcolor}

\begin{document}

\title{Nonreciprocal Chiral Automata}

\author{Andrew A. Allocca}
\affiliation{Physics Department, City College of the City University of New York, New York 10031, USA}

\author{Armin Rahmani}
\affiliation{Department of Physics and Astronomy and Advanced Materials Science and Engineering Center, Western Washington University, Bellingham, Washington 98225, USA}

\author{Pouyan Ghaemi}
\affiliation{Physics Department, City College of the City University of New York, New York 10031, USA}
\affiliation{Physics Program, Graduate Center of City University of New York, New York 10031, USA}

\author{Sriram Ganeshan}
\affiliation{Physics Department, City College of the City University of New York, New York 10031, USA}
\affiliation{Physics Program, Graduate Center of City University of New York, New York 10031, USA}

\begin{abstract}
Odd pressure is a parity-odd transport coefficient that gives an isotropic pressure response to vorticity in compressible two-dimensional fluids. By asymmetrically coupling vortical and compressional modes, it provides a direct mechanism for a nonreciprocal hydrodynamic response reminiscent of several biological systems. Yet microscopic models realizing odd pressure as a transport coefficient are relatively scarce. Here, we construct one such model using a lattice-gas cellular automaton based on the Frisch-Hasslacher-Pomeau II model with rest particles and chiral collisions, whose Chapman-Enskog coarse-graining produces a nonzero odd pressure coefficient. The key ingredient is a local parity-breaking collision that weakly rotates moving particles in the presence of a rest particle. At hydrodynamic scales, this microscopic rule acts as an effective magnetic field, converting the model's bulk viscosity into odd pressure. More generally, we derive expressions relating Hall viscosity, odd pressure, and odd torque to their parity-even counterparts in our model. 
\end{abstract}

\maketitle

\textit{Introduction.---}Broken parity is a widespread feature of systems whose microscopic constituents acquire handedness through geometry, internal rotation, dynamics, or coupling to external fields. 
At long wavelengths, this microscopic chirality can appear as parity-odd transport. 
At fixed chirality, such transport can generate nonreciprocity, whereby exchanging the driven and measured hydrodynamic channels changes the response \cite{You2020,Fruchart2021}. 
Nonreciprocal couplings occur in systems ranging from active mixtures and flocks to driven interfaces and photonic devices \cite{Dadhichi2020,AgudoCanalejo2019,Pan1994,Metelmann2015,Hosaka2023}. 
Microscopically, effective nonreciprocal interactions often reflect broken action--reaction symmetry within a reduced subsystem \cite{Ivlev2015}; hydrodynamically, they are encoded in asymmetric cross-couplings between collective modes.

\begin{figure}[!ht]
    \centering
    \includegraphics[width=0.93\linewidth]{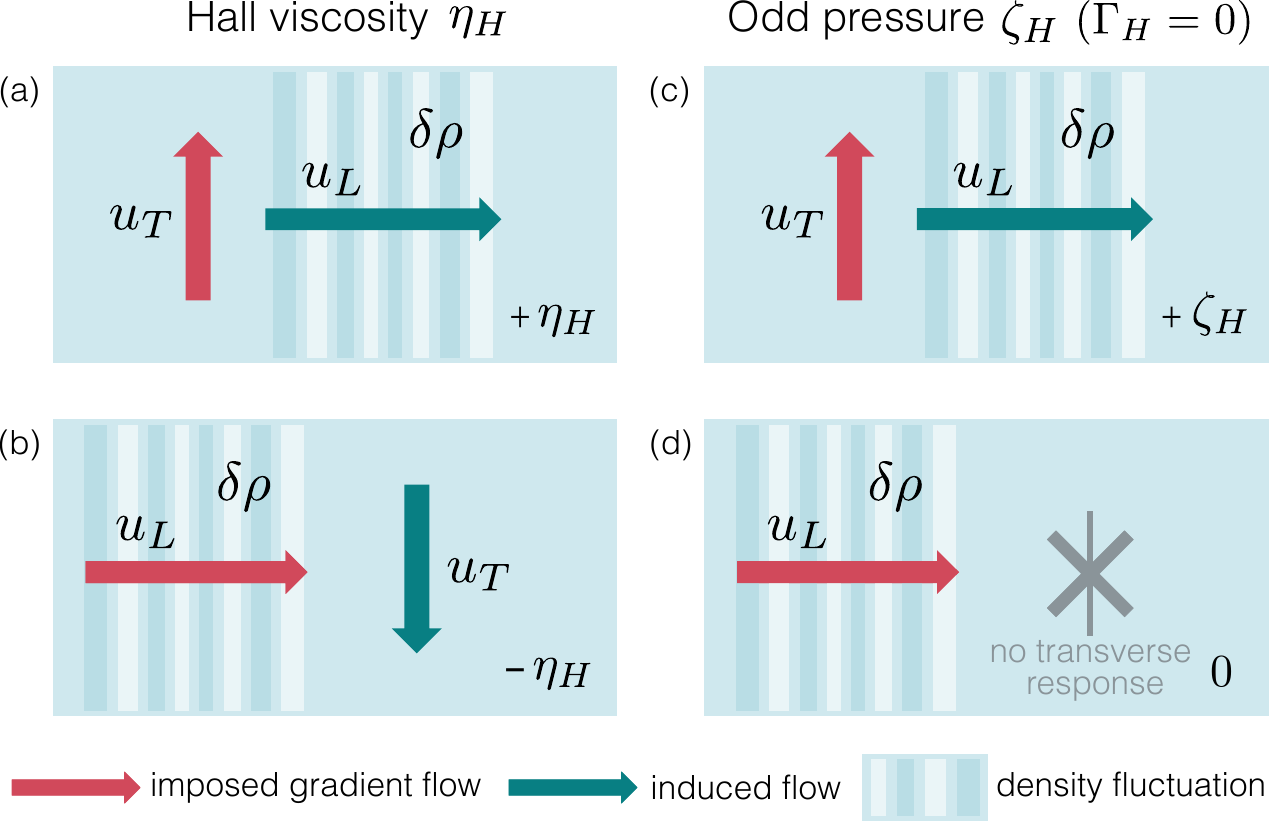}
    \caption{The nonreciprocal effect of odd pressure $\zeta_H$ can be seen in the response to imposed nonuniform flows, as in \cref{eq:linear-response}. 
    Hall viscosity generates both (a) longitudinal flow $u_L$ from transverse flow gradients (b) transverse flows $u_T$ from longitudinal flow gradients. 
    Odd pressure $\zeta_H$ generates (c) longitudinal flows from transverse flow gradients, but (d) there is no effect from longitudinal flow gradients---a nonreciprocal response.
    In all cases changing $u_L$ induces density fluctuations $\delta\rho$.}
    \label{fig:response}
\end{figure}

The most familiar parity-odd transport coefficient is odd, or Hall, viscosity, a nondissipative transverse viscous response studied in quantum Hall fluids, plasmas, electron hydrodynamics, active matter, and classical chiral fluids \cite{Avron1995, Ganeshan2017, Banerjee2017, Soni2019, Scheibner2020, Fruchart2023, Nassar2020, Hargus2021, Srivastava2024, Qi2023, Jiao2026a, Jiao2026b, Lier2026, Maire2026, Tokatly2006, Tokatly2007, Haldane2011, Hoyos2012, Bradlyn2012, Abanov2013, Hoyos2014, Laskin2015, Can2015, Klevtsov2015, Scaffidi2017, Pellegrino2017, Berdyugin2019, Korving1966, Markovich2021, Reynolds2022}. 
Compressible isotropic fluids, however, support additional parity-odd responses, including odd pressure and odd torque \cite{Monteiro2023}. 
Odd pressure can render the linear hydrodynamic response nonreciprocal, as demonstrated schematically in \cref{fig:response}. Markovich and Lubensky obtained analogous nonreciprocal momentum dynamics by coarse-graining a general molecular description of externally torqued chiral active fluids with driven angular-momentum density \cite{Markovich2024Nonreciprocity}; however, a concrete microscopic mechanism realizing this response remained lacking. More broadly, microscopic chirality does not necessarily survive coarse-graining, and the mechanisms that select specific macroscopic odd transport coefficients remain poorly understood in compressible fluids, particularly for odd pressure. 

Several complementary microscopic approaches have explored how macroscopic parity-odd transport, particularly odd viscosity, emerges from microscopic chirality.
Kinetic theories and active-matter models of chiral rotors showed how spinning constituents and frictional collisions convert internal angular momentum into odd viscosity \cite{Banerjee2017}.
Colloidal spinner experiments and particle simulations then demonstrated odd-fluid phenomenology in explicitly microscopic media \cite{Soni2019, Han2021}.

Recent work on chirality arising from collisions and geometry has been systematically coarse-grained using Boltzmann kinetic theory, yielding explicit expressions for the associated odd transport coefficients~\cite{Han2021, Fruchart2022, Eren2025}.
Statistical-mechanical treatments derive Green-Kubo relations and Hamiltonian microscopic mechanisms for odd viscosity in fluids with internal spin or aligned angular momentum \cite{Epstein2020,Markovich2021}. 
In recent work, we pursued this bottom-up approach in a discrete, analytically tractable setting using a chiral Frisch–Hasslacher–Pomeau (FHP) lattice-gas automaton, in which parity-breaking two-body collisions and weak microscopic velocity rotations generate Hall viscosity upon Chapman–Enskog coarse-graining \cite{Allocca2026}.

Here we show that the FHP-II extension of the FHP gas, which includes a non-mobile rest state \cite{Frisch1986}, provides the missing ingredient for a microscopic realization of odd pressure and the associated macroscopic nonreciprocal response. 
The rest state endows the coarse-grained fluid with a nonzero bulk viscosity $\zeta$. 
We then introduce a local parity-breaking collision rule so that when a rest particle is present, the moving particles undergo a weak chiral rotation. 
At hydrodynamic scales, this rule acts as an effective magnetic field and converts $\zeta$ into a finite odd-pressure coefficient $\zeta_H$. 
More generally, we derive relations showing how the parity-odd coefficients can be inherited from their corresponding even-parity coefficients.

\textit{Hydrodynamic odd pressure.—} The viscosity tensor $\eta_{ijkm}$ is defined through the relation $\sigma_{ij}=\eta_{ijkm}\partial_k u_m$, where $\sigma_{ij}$ is the viscous stress tensor and $\partial_k u_m$ is the velocity gradient.
In a Galilean-invariant first-order hydrodynamic theory, the viscosity tensor contains six independent coefficients \cite{Monteiro2023},
\begin{multline}
    \eta_{ijkm}
    = \eta(\delta_{ik}\delta_{jm}+\delta_{im}\delta_{jk}-\delta_{ij}\delta_{km})
    +\zeta\,\delta_{ij}\delta_{km}
    +\Gamma\,\epsilon_{ij}\epsilon_{km} \\
    +\eta_H(\epsilon_{ik}\delta_{jm}+\epsilon_{jm}\delta_{ik})
    +\zeta_H\,\delta_{ij}\epsilon_{km}
    +\Gamma_H\,\epsilon_{ij}\delta_{km}.
    \label{eq:viscosity-tensor}
\end{multline}
The coefficients $\eta$, $\zeta$, and $\Gamma$ are the shear, bulk, and rotational viscosities, while the parity-odd coefficients are the Hall viscosity $\eta_H$, the odd pressure $\zeta_H$, and the odd torque $\Gamma_H$.
The coefficient $\eta_H$ produces shear stresses transverse to strain.
By contrast, $\zeta_H$ gives an isotropic pressure response to vorticity and is therefore intrinsically tied to compressibility.

Before turning to the microscopic dynamics underlying odd pressure, we first review the key distinction between odd-viscous and odd-pressure responses in linearized compressible hydrodynamics. 
Being odd, both $\eta_H$ and $\zeta_H$ coefficients connect fluid flow in one direction to gradients of the flow in the transverse direction, but while $\eta_H$ does this in a completely antisymmetric way, the odd pressure $\zeta_H$ may break this relationship and generate nonreciprocal behavior, as we show below and in \cref{fig:response}---a gradient in the transverse velocity of a hydrodynamic mode produces flow along the direction of the gradient, but not vice versa. 
Despite this simple hydrodynamic interpretation, odd pressure has received far less attention from a microscopic perspective than Hall viscosity, and the chiral FHP automaton of Ref.~\cite{Allocca2026} still had $\zeta_H=0$.

Consider a mode with wave vector along $\hat{x}$; write $u_L=u_x$ (longitudinal) and $u_T=u_y$ (transverse), and linearize about a uniform state of density $\rho_0$, with $c_s^2=\eval{(\partial P/\partial\rho)}_{\rho=\rho_0}$ the squared speed of sound and pressure $P$.
Setting the rotational viscosity $\Gamma$ and odd-torque $\Gamma_H$ to zero, as is the case for our automaton model, continuity and momentum conservation give
\begin{align}
    \partial_t\delta\rho={}&-\rho_0\partial_xu_L, \notag\\
    \rho_0\partial_tu_a={}&-\delta_{aL}c_s^2\partial_x\delta\rho
    +\rho_0\mathsf{D}_{ab}\partial_x^2u_b, \notag\\
    \mathsf{D}={}&
    \begin{pmatrix}
        \eta+\zeta & \eta_H+\zeta_H\\
        -\eta_H & \eta
    \end{pmatrix},
    \label{eq:linear-response}
\end{align}
where $a,b\in\{L,T\}$ and $\delta\rho$ is the small deviation of the density from $\rho_0$.
Hall viscosity therefore relates longitudinal and transverse motion with equal magnitude and opposite sign, whereas odd pressure contributes only to the transverse-to-longitudinal response, shown schematically in \cref{fig:response}; the mismatch between the two cross responses is precisely $\zeta_H$.
This nonreciprocity \footnote{In the full six-coefficient theory~\cref{eq:viscosity-tensor}, $\zeta_H$ is Onsager-paired with the odd-torque coefficient $\Gamma_H$, which can supply the reverse longitudinal-to-transverse response \cite{Monteiro2023}.
Thus, the nonreciprocity here is not a property of odd pressure alone, but of the $\Gamma_H=0$ sector realized by our automaton.} is observable only when the longitudinal density mode is dynamical.
In the incompressible limit $\partial_i u_i=0$, one has $u_L=0$ at finite wave vector, and the odd-pressure force $\partial_i(\zeta_H\epsilon_{km}\partial_k u_m)$ can be absorbed into the Lagrange-multiplier pressure, $P_{\mathrm{eff}}=P-\zeta_H\epsilon_{km}\partial_k u_m$.
For a compressible fluid, by contrast, the equation of state ties pressure to density, so transverse vorticity can excite a distinct longitudinal density and sound response.
Related nonreciprocal hydrodynamic momentum dynamics were derived microscopically for chiral active fluids with driven angular-momentum density by Markovich and Lubensky \cite{Markovich2024Nonreciprocity}.
%Experimentally, weakly compressible magnetic colloidal-rotor monolayers already permit simultaneous imaging of velocity, vorticity, and density \cite{Mecke2023ActiveTurbulence}, suggesting a direct test in which a transverse modulation is applied and the induced longitudinal density wave is measured. 
We now develop the microscopic automaton model that gives rise to odd pressure and identify the specific features of the microscopic dynamics responsible for the nonreciprocity discussed above.

\textit{From FHP-I to chiral FHP-II.---}
We first review the structure of the FHP-I lattice gas and its chiral generalization in our previous work~\cite{Allocca2026}. 
The model lives on a triangular lattice and assigns six Boolean occupation variables $N_\ell\in\{0,1\}$ with $\ell=1,\dots,6$ to each site, corresponding to the six velocities $\hat{\mathbf{c}}_\ell = \big(\cos(\pi(\ell-1)/3),\sin(\pi(\ell-1)/3)\big)$, where $\ell$ is defined modulo six. 
The update consists of a local collision step followed by streaming along the outgoing velocity links.
The allowed FHP-I collisions conserve both particle number and momentum. 
The elementary two-body process is the scattering of a head-on pair into either of the two other head-on pairs related by a clockwise or counter-clockwise rotation; the model also includes the usual three-body collision between the two alternating triplets of links.
When the two possible two-body scattering outcomes occur with equal probability, the model preserves parity symmetry; biasing the outcomes, giving a clockwise rotation probability $p$ and counterclockwise probability $1-p$, produces a chiral FHP-I model.
This gives a microscopic handedness while preserving the number- and momentum-conserving structure of the automaton, a key ingredient to produce Hall viscosity in the FHP-I analysis.

\begin{figure*}[!ht]
    \centering
    \includegraphics[width=0.8\linewidth]{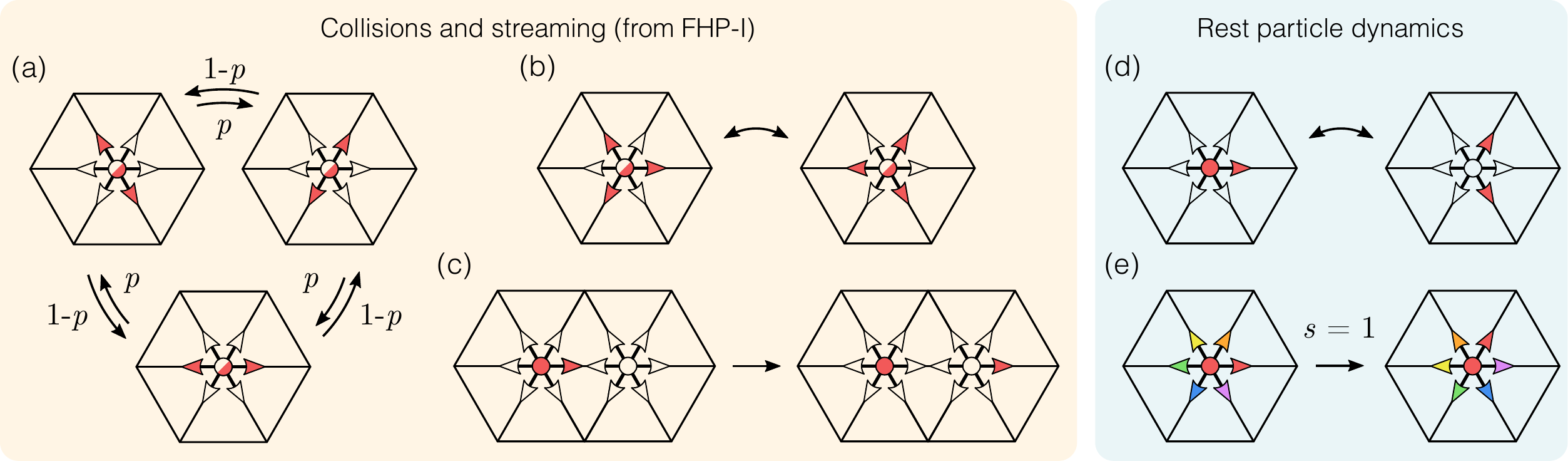}
    \caption{Schematic depictions of the collision and streaming rules for the FHP-II automaton, separated into processes inherited from FHP-I and new dynamics of rest particles. Arrows represent mobile states and the central circle the rest state. The two-body and three-body collisions are shown in (a) and (b), with a half-filled circle indicating insensitivity to rest particles. Streaming is shown in (c): mobile particles move to the next vertex along their velocity and rest particles remain in place. The collisions that change the rest particle occupation are shown in (d), and in (e) we show the rotation of all mobile state occupations when a rest particle is present.}
    \label{fig:automatonrules}
\end{figure*}

The FHP-II model is obtained by adding one more Boolean state at each site, $N_0\in\{0,1\}$ with velocity $\hat{\mathbf c}_0=0$---a rest particle that may participate in collision but does not propagate, unlike the six mobile states ($\ell\neq0$) inherited from FHP-I. 
This seven-bit extension was introduced in the early lattice-gas literature as a practical improvement of the original FHP-I model, increasing the available collision channels by allowing particles to be temporarily stored in a nonpropagating state \cite{FHPReview1987,dHumieres1987}.
One important hydrodynamic consequence, already identified in those analyses, is that while the original six-bit FHP-I model has vanishing bulk viscosity, the rest-particle collisions of FHP-II produce a nonzero bulk viscosity. 
This feature is central for us, since odd pressure will arise by mixing this bulk-viscous channel with a parity-breaking microscopic rotation. 

The automaton rules, including rest particles, are given in \cref{fig:automatonrules}. 
The streaming rule and the two- and three-body collisions between mobile particles from FHP-I are unchanged by the presence or absence of a rest particle spectator, shown in \cref{fig:automatonrules}(a,b,c).
Collisions which allow scattering of particles into and out of the rest state while conserving particle number and momentum are shown in \cref{fig:automatonrules}(d): a pair of mobile particles in states $\ell+1$ and $\ell-1$ scatter to a mobile particle in state $\ell$ and a rest particle, or vice versa.
This collision process also preserves momentum. 

Finally, we add a new microscopic parity-breaking rule: when an initial scattering state contains a rest particle, all mobile particle velocities are rotated chirally, $N_\ell\to N_{\ell+s}$ for $\ell\neq0$, shown in \cref{fig:automatonrules}(e) for $s=1$. 
In Ref.~\cite{Allocca2026} we considered the effect of a similar velocity rotation applied uniformly to all particles, which we showed gives the Lorentz force exerted by a uniform external magnetic field upon coarse-graining. 
The rest-particle-induced rotation here can be interpreted as the effect of a local magnetic moment, or spin, carried by rest particles themselves, and represents a new microscopic parity-breaking effect that the simpler FHP-I model cannot support. 
The inclusion of this new rest-particle rule is what will generate odd pressure $\zeta_H$ in the long-wavelength hydrodynamic model.
Unlike parity-breaking via chiral scattering, this rotation of particle velocities changes momentum at a vertex, a key ingredient for introducing nonreciprocal effects.

Denoting the ensemble-averaged occupations by $n_\ell=\langle N_\ell\rangle$, the Boltzmann equation incorporating all of these new scattering and rotational effects is 
\begin{equation}
    n_{\ell+sn_0}(\mathbf{x}+\hat{\mathbf{c}}_\ell,t+1)-n_\ell(\mathbf{x},t) = \Omega^{II}_\ell[\left\{n\right\}],
\end{equation}
where we have taken the initial rest-particle rotation of each time step and moved it to the end of the previous time step; this is easier to treat analytically. 
The FHP-II collision function $\Omega^{II}_\ell$ can be written in terms of the chiral FHP-I collision function $\Omega_\ell^I$ supplemented by collisions into and out of the rest state, and is given explicitly in the End Matter.
Conservation of particle number and momentum give $\sum_\ell\Omega^{II}_\ell =0$ and $\sum_\ell\hat{\mathbf{c}}_\ell \Omega^{II}_\ell = 0$.

The addition of the rest-particle state necessarily changes the equilibrium solution of link occupation, which is the starting point for developing the macroscopic hydrodynamic theory of this automaton.
The local particle density and momentum of the model are $\rho=\sum_{\ell=0}^6 n_\ell(\mathbf{x},t)$ and $\rho\mathbf{u}=\sum_{\ell=0}^6\hat{\mathbf c}_\ell n_\ell$, with the latter written with the local velocity field $\mathbf{u}$.
When these vary slowly in space and time, in the low Mach number regime $\abs{\mathbf{u}}\ll1$ the equilibrium occupation is
\begin{equation}
    n^\mathrm{eq}_\ell\approx \frac{\rho}{7} + \frac{\rho}{3}\hat{c}_\ell^i u_i+\rho\, G(\rho)Q^{ij}_{\ell}u_i u_j,
\end{equation}
which has a similar form as for FHP-I, but with key differences. 
The lowest-order term is $\rho/7$ since there are 7 states at each vertex, while for FHP-I this term is $\rho/6$. 
We also have $G(\rho) = \tfrac{7}{18}\tfrac{7-2\rho}{2-\rho}$, and $Q^{ij}_{\ell}\equiv\hat{c}_\ell^i\hat{c}_\ell^j-\frac{3}{7}\delta^{ij}$ is the traceless rank-2 tensor, with $\sum_{\ell=0}^6Q^{ij}_{\ell}=0$.
The differences from FHP-I reflect a change to the speed of sound from $c_s=1/\sqrt{2}$ to the FHP-II value $c_s=\sqrt{3/7}$.

\textit{Hydrodynamic theory.---}The Chapman-Enskog expansion of the Boltzmann equation assumes $\rho$ and $\mathbf{u}$ to vary on long length scales, so that $n_\ell$ is near the equilibrium solution in large patches of size $1/\varepsilon \gg 1$ in terms of the lattice constant.
Likewise we assume these evolve slowly in time, with sound waves propagating on scale $\sim1/\varepsilon$ and advective and diffusive phenomena on yet slower scale $\sim1/\varepsilon^2$.
We also consider the rest-particle-induced rotation to be weak, so that the resultant dynamics occur at these same large length and slow time scales. 
This can be implemented by generalizing the discrete directional index $\ell$ to a continuous angular variable $\theta$ that is gradually incremented by the spin $S\sim\epsilon$ of rest particles whenever they are present.
The Chapman-Enskog expansion including the weakened rotational effect is discussed in detail in Ref.~\cite{Allocca2026}. 

By expanding the Boltzmann equation up to $O(\varepsilon^2)$ we obtain the hydrodynamic equations
\begin{gather}
    \partial_t\rho+\partial_i(\rho \tilde{u}_i)=0 \label{eq:continuity}\\
    \partial_t\left(\rho \tilde{u}_i\right)= -\mathcal{S}\,\epsilon^{ij}\left(\rho \tilde{u}_j\right) - S\, \varepsilon n_0^1\,\epsilon^{ij}(\rho u_j) + \partial_j\sigma_{ij},\label{eq:NavierStokes}
\end{gather}
where $\mathcal{S}=S\rho/7$ is the average spin density of the rest particles and $\rho\tilde{u}_i=(\delta^{ij}+\tfrac{1}{2}\mathcal{S}\epsilon^{ij})\rho u_j$.
The average spin density acts as a uniform magnetic field and enters the Navier-Stokes equation \ref{eq:NavierStokes} in a Lorentz force term, causing the transverse deflection of the fluid momentum reflected in $\rho\tilde{u}_i$. 
Since this $\rho\tilde{u}_i$ is what appears in the continuity equation \ref{eq:continuity}, it is the physical notion of local fluid momentum in the system. 
The fluctuations of the rest particle occupation $\varepsilon\,n_0^1$ cause a fluctuation in the average spin density, reflected as second Lorentz force term in \cref{eq:NavierStokes} that is not present when considering a uniform external field. 
The stress tensor
\begin{multline}
    \sigma_{ij} = -\frac{3}{7}\rho\,\delta^{ij}-\rho\,G(\rho)\sum_\ell\hat{c}^i_\ell\hat{c}^j_\ell Q^{km}_\ell u_k u_m\\
    -\sum_{\ell}\hat{c}_\ell^i\hat{c}_\ell^j\left[\frac{1}{6}Q^{km}_{\ell}\partial_k(\rho u_m)+\varepsilon \,n_\ell^1\right],
\end{multline}
depends explicitly on the small fluctuations of the mobile particle occupations from local equilibrium $\varepsilon\,n_\ell^1$ in the viscous term.

The expansion gives $\sigma_{ij}$ in terms of $\rho u_i$ instead of $\rho\tilde{u}_i$, missing the magnetic effect of the rest particle spin density. 
Since the kinematic viscosity tensor parametrizes the stress contribution from gradients of the fluid's physical momentum, we must substitute $\rho u_i$ in terms of $\rho\tilde{u}_i$ to obtain the full viscous effect. 
As a result, we find that the viscosity with the effect of $\mathcal{S}$ can be written in terms of the viscosity for $\mathcal{S}=0$,
\begin{gather}
    \sigma_{ij} = -P\delta^{ij}+\eta_{ijkm}^0\partial_k(\rho u_m) = -P\delta^{ij}+\eta_{ijkm}\partial_k(\rho\tilde{u}_m) \nonumber\\
    \Rightarrow \eta_{ijkm} = \eta_{ijkn}^0\frac{\delta^{nm}-\tfrac{\mathcal{S}}{2}\epsilon^{nm}}{1+\mathcal{S}^2/4},
\end{gather}
where $\eta_{ijkm}^0\equiv \eval{\eta_{ijkm}}_{\mathcal{S}=0}$.
We therefore only need microscopic calculations for the fluctuations $\varepsilon\,n^1_\ell$, from which we obtain the viscosity, for $\mathcal{S}=0$. 

\textit{Fluctuations and viscosity.---} The general machinery we developed in Ref.~\cite{Allocca2026} to compute fluctuations applies here without modification, so that for all $\ell$
\begin{equation} \label{eq:n1general}
    \varepsilon\,n^1_\ell = \sum_{\ell'}\sum_{\alpha\vert\lambda_\alpha\neq 0}\frac{1}{3\lambda_\alpha}v^\alpha_\ell(v^\alpha_{\ell'})^\ast Q^{ij}_{\ell'}\partial_i(\rho u_j),
\end{equation}
where $\lambda_\alpha$ and $\mathbf{v}^\alpha$ are the eigenvalues and eigenvectors of the linearized collision matrix $\Lambda^{II}_{\ell\ell'} = \partial\Omega^{II}_\ell/\partial n_{\ell'}\vert_{n=n^0}$.
Three of the eigenvectors of $\Lambda^{II}$ have zero eigenvalues, reflecting conservation of particle number and the two components of momentum; the other four have nonzero eigenvalues and may contribute in \cref{eq:n1general}.
The form of $\Lambda^{II}$, its eigenvalues and eigenvectors, and $\varepsilon\,n^1_\ell$ are all given in the End Matter. 

The fluctuation of the rest particle occupation $n^1_0$ does not contribute to the viscosity, though it does appear in \cref{eq:NavierStokes} capturing the contribution to the Lorentz force from fluctuations of the average spin density. 
With $n_{\ell\neq0}^1$ we compute the $\mathcal{S}=0$ viscosity coefficients for the FHP-II model with chiral two-body scattering, 
\begin{gather}
    \eta^0 = \frac{1}{4}\frac{3\gamma+4\kappa}{(3\gamma+4\kappa)^2+12\gamma^2\left(p-\tfrac{1}{2}\right)^2}-\frac{1}{8} \label{eq:eta0}\\
    \eta_H^0 = -\frac{\sqrt{3}}{2}\frac{\gamma\left(p-\tfrac{1}{2}\right)}{(3\gamma+4\kappa)^2+12\gamma^2\left(p-\tfrac{1}{2}\right)^2} \label{eq:etaH0}\\
    \zeta^0 = \frac{1}{98\kappa}-\frac{1}{28} \label{eq:zeta0},
    % \eta^0 = \frac{1}{24\gamma} \frac{\frac{3}{2} + 2(1-\rho/7)}{\frac{3}{4}+2(1-\rho/7)+\frac{4}{3}(1-\rho/7)^2+(p-\tfrac{1}{2})^2}-\frac{1}{8}\\
    % \eta_H^0 = \frac{\sqrt{3}}{24\gamma}\frac{p-\tfrac{1}{2}}{\frac{3}{4}+2(1-\rho/7)+\frac{4}{3}(1-\rho/7)^2+(p-\tfrac{1}{2})^2}\\
    % \zeta^0 = \frac{1}{98\kappa}-\frac{1}{28}\\
\end{gather}
and $\zeta_H^0=\Gamma^0=\Gamma_H^0=0$, where $\gamma = (\rho/7)(1-\rho/7)^3$, $\beta = (\rho/7)^2(1-\rho/7)^2$, and $\kappa=(\rho/7)(1-\rho/7)^4$.
The result for the $\mathcal{S}=0$ shear viscosity $\eta^0$ reduces to the known result for the parity-symmetric FHP-II model in the limit $p=1/2$, and the bulk viscosity $\zeta^0$ is the known result~\cite{FHPReview1987,dHumieres1987}, unchanged by the inclusion of chiral two-body scattering.
The Hall viscosity $\eta_H^0$ is similar to the form obtained for the FHP-I model~\cite{Allocca2026}. 
The parity-breaking effects of chiral collisions alone are insufficient to generate any other viscosity coefficients. 

As discussed above, the viscosity components including the rotation by rest particle spin density $\mathcal{S} = S\rho/7$ are obtained from these after substitution of the physical momentum $\rho\tilde{u}_i$,
\begin{gather}
    \eta = \frac{\eta^0 +\tfrac{\mathcal{S}}{2} \eta_H^0}{1+\mathcal{S}^2/4} \quad \eta_H = \frac{\eta_H^0 - \tfrac{\mathcal{S}}{2}\eta^0}{1+\mathcal{S}^2/4} \label{eq:magneticeta}\\
    \zeta = \frac{\zeta^0 +\tfrac{\mathcal{S}}{2} \zeta_H^0}{1+\mathcal{S}^2/4} = \frac{\zeta^0}{1+\mathcal{S}^2/4} \label{eq:zeta}\\
    \zeta_H = \frac{\zeta_H^0 - \tfrac{\mathcal{S}}{2}\zeta^0}{1+\mathcal{S}^2/4} = -\frac{\tfrac{\mathcal{S}}{2}\zeta^0}{1+\mathcal{S}^2/4} \label{eq:zetaH}. 
\end{gather}
where the first expression in \cref{eq:zeta,eq:zetaH} are generic and the second substitutes $\zeta_H^0=0$. 
The $\mathcal{S}=0$ shear and Hall viscosities $\eta^0$ and $\eta_H^0$ mix to give the full $\eta$ and $\eta_H$, as found in Ref.~\cite{Allocca2026} for the FHP-I automaton with uniform external magnetic field.
We now also find that a similar relationship holds for $\zeta$ and $\zeta_H$ with their respective zero-spin counterparts.
Unlike for the Hall viscosity $\eta_H$, which results from either parity-breaking mechanism considered, chiral collisions alone are not sufficient to produce an odd pressure $\zeta_H$ in these automaton models, and a parity-breaking magnetic effect is necessary to generate $\zeta_H$ from the FHP-II model's bulk viscosity. 
Though here we include this via a local rotation induced by the rest particle spin-density, the macroscopic effect has the form of an external force which changes the total momentum of the system.
This is ultimately how the microscopic dynamics we have defined allow for the nonreciprocal effects induced by $\zeta_H$ as shown in Fig.~\ref{fig:response} and outlined in Eq.~\ref{eq:linear-response}.

\textit{Conclusion and further directions.} We have demonstrated how parity-breaking extensions to the FHP-II lattice automaton generate odd-parity viscosity coefficients in the hydrodynamic theory obtained through coarse-graining. 
We find that the odd pressure coefficient $\zeta_H$, giving nonreciprocal hydrodynamic effects, does not result from parity-breaking in collisions alone, and a magnetic effect is necessary to generate this term from the automaton's bulk viscosity $\zeta$, included here as the chiral deflection of mobile particles by rest particles. 
Because of this insensitivity to the parity breaking of collisions, the two odd viscosities $\eta_H$ and $\zeta_H$ are independent in this model, so that the collision parameter $p$ can in principle be tuned to make $\eta_H$ vanish while $\zeta_H$ remains finite. 

The only remaining viscosity coefficients yet to be found in an automaton model are the rotational viscosity $\Gamma$ and the odd torque $\Gamma_H$. 
In the same way that \cref{eq:magneticeta,eq:zeta,eq:zetaH} relate $\eta,\eta_H,\zeta$, and $\zeta_H$ to their non-magnetic counterparts, a similar relationship is found for two coefficients,
\begin{equation}
    \Gamma = \frac{\Gamma^0-\tfrac{\mathcal{S}}{2}\Gamma_H^0}{1+\mathcal{S}^2/4} \quad \Gamma_H = \frac{\Gamma_H^0+\tfrac{\mathcal{S}}{2}\Gamma^0}{1+\mathcal{S}^2/4},
\end{equation}
with the same $\mathcal{S}$ as above. 
Therefore, any microscopic automaton rule that generates one is sufficient to generate the other when supplemented with a magnetic effect such as the spin-density-generated internal field as considered here or an external field. 
Such a rule in an FHP-type automaton must be more complex than a local scattering rule, however, since $\Gamma$ and $\Gamma_H$ correspond to processes involving a notion of local angular momentum which is not present in these models. 
Extensions such as the inclusion of global, or at least nonlocal, scattering processes or adding new local degrees of freedom carried by the automaton particles may lead to realization of these additional coefficients.

\begin{acknowledgments}
\textit{Acknowledgements.---}This work was supported by the National Science Foundation (NSF) Grant No.~DMR-2315063 (A.A., P.G., and S.G.) and NSF Grant No.~DMR-2315064 (A.R.).  S.G. was supported in part by grant NSF PHY-2309135 to the Kavli Institute for Theoretical Physics (KITP), where part of this work was carried out and would also like to thank Kranthi K. Mandadapu for useful discussions and sharing related unpublished results. 
\end{acknowledgments}

\bibliography{references.bib}

\appendix

\onecolumngrid
\section{{\large End Matter}} 

The collision function of the FHP-II model is
\begin{multline}
    \Omega_\ell^{II} = (1-\delta_{\ell0})\bigg\{\Omega_\ell^{I} +\left[(1-n_0)(1-n_\ell)n_{\ell+1}n_{\ell-1}-n_0 n_\ell(1-n_{\ell+1})(1-n_{\ell-1})\right](1-n_{\ell+2})(1-n_{\ell-2})(1-n_{\ell+3})\\
    +n_0 (1-n_\ell)\left[n_{\ell+1}(1-n_{\ell-1})+n_{\ell-1}(1-n_{\ell+1})\right](1-n_{\ell+2})(1-n_{\ell-2})(1-n_{\ell+3})\\
    -(1-n_0)n_\ell\left[n_{\ell+2}(1-n_{\ell-2})+n_{\ell-2}(1-n_{\ell+2})\right](1-n_{\ell+1})(1-n_{\ell-1})(1-n_{\ell+3})\bigg\} \\
    +\delta_{\ell0}\left[(1-n_0)\sum_{k=1}^6 n_{k+1} n_{k-1} (1-n_k)(1-n_{k+2})(1-n_{k-2})(1-n_{k+3})\right. \\
    \left.+n_0\sum_{k=1}^6n_k (1-n_{k+1})(1-n_{k-1})(1-n_{k+2})(1-n_{k-2})(1-n_{k+3})\right].
\end{multline}
where $\Omega_\ell^I$ is the collision function of the chiral FHP-I model given in \cite{Allocca2026}. 
Substituting $n_\ell = n_\ell^0+\varepsilon n_\ell^1$, expanding to linear order in $\varepsilon$, and evaluating at $\mathbf{u}=0$ gives the linearized collision function for the chiral FHP-II model
\begin{equation}
    \Lambda^{II} = 
    \begin{pmatrix}
        0 & \mathbf{0}^T \\
        \mathbf{0} & \Lambda^{I}
    \end{pmatrix}
    +\kappa
    \begin{pmatrix}
        -6 & 1 & 1 & 1 & 1 & 1 & 1 \\
        1 & -3 & 2 & -1 & 0 & -1 & 2 \\
        1 & 2 & -3 & 2 & -1 & 0 & -1 \\
        1 & -1 & 2 & -3 & 2 & -1 & 0 \\
        1 & 0 & -1 & 2 & -3 & 2 & -1 \\
        1 & -1 & 0 & -1 & 2 & -3 & 2 \\
        1 & 2 & -1 & 0 & -1 & 2 & -3 \\
    \end{pmatrix},
\end{equation}
where $\Lambda^I$ is the FHP-I linearized collision matrix
\begin{equation}
    \Lambda^{I} = \begin{pmatrix} 
        -\gamma-\beta & \gamma(1-p)+\beta & p\gamma-\beta & -\gamma+\beta & \gamma(1-p)-\beta & p\gamma+\beta \\
        p\gamma+\beta & -\gamma-\beta & \gamma(1-p)+\beta & p\gamma-\beta & -\gamma+\beta & \gamma(1-p)-\beta \\
        \gamma(1-p)-\beta & p\gamma+\beta & -\gamma-\beta & \gamma(1-p)+\beta & p\gamma-\beta & -\gamma+\beta \\
        -\gamma+\beta & \gamma(1-p)-\beta & p\gamma+\beta & -\gamma-\beta & \gamma(1-p)+\beta & p\gamma-\beta \\
        p\gamma-\beta & -\gamma+\beta & \gamma(1-p)-\beta & p\gamma+\beta & -\gamma-\beta & \gamma(1-p)+\beta \\
        \gamma(1-p)+\beta & p\gamma-\beta & -\gamma+\beta & \gamma(1-p)-\beta & p\gamma+\beta & -\gamma-\beta 
    \end{pmatrix}.
\end{equation}
These matrices are written in terms of the constants 
\begin{gather*}
    \gamma = \eval{n_\ell^0\left(1-n_\ell^0\right)^3}_{\mathbf{u}=0} = (\rho/7)(1-\rho/7)^3 \\
    \beta = \eval{\left(n_\ell^0\right)^2\left(1-n_\ell^0\right)^2}_{\mathbf{u}=0} = (\rho/7)^2(1-\rho/7)^2 \\
    \kappa=\eval{n_\ell^0\left(1-n_\ell^0\right)^4}_{\mathbf{u}=0}=(\rho/7)(1-\rho/7)^4.
\end{gather*}
The seven eigenvectors and eigenvalues of $\Lambda^{II}$ are
\begin{align*}
    \mathbf{v}^1 &= (-6,1,1,1,1,1,1)/\sqrt{42} &  & \lambda_1 = -7\kappa\\
    \mathbf{v}^2 &= (0,1,e^{2\pi i/3},e^{-2\pi i/3},1,e^{2\pi i/3},e^{-2\pi i/3})/\sqrt{6} & & \lambda_2 = -3\gamma-4\kappa-i2\sqrt{3}\gamma\left(p-\tfrac{1}{2}\right) \\
    \mathbf{v}^3 &= (0,1,-1,1,-1,1,-1)/\sqrt{6} & & \lambda_3 = -3(2\beta+3\kappa)\\
    \mathbf{v}^4 &= (0,1,e^{-2\pi i/3},e^{2\pi i/3},1,e^{-2\pi i/3},e^{2\pi i/3})/\sqrt{6} & & \lambda_4 = -3\gamma-4\kappa+i 2\sqrt{3}\gamma\left(p-\tfrac{1}{2}\right) \\
    \mathbf{v}^5 &= (1,1,1,1,1,1,1)/\sqrt{7} & & \lambda_5 = 0\\
    \mathbf{v}^6 &= (0,1,\tfrac{1}{2},-\tfrac{1}{2},-1,-\tfrac{1}{2},\tfrac{1}{2})/\sqrt{3} & & \lambda_6 = 0\\
    \mathbf{v}^7 &= (0,0,1,1,0,-1,-1)/2, & & \lambda_7 = 0.
\end{align*}
The three with vanishing eigenvalues correspond to exactly conserved quantities---density and 2d momentum. 
Since the real part of all non-vanishing eigenvalues is negative, the remaining eigenvectors correspond to dissipative modes and thus contribute to fluctuations through \cref{eq:n1general}.
Explicitly we find that the fluctuations of the FHP-II model are 
\begin{gather}
    \varepsilon\, n_0^1 = \frac{\delta^{ij}}{49\kappa}\partial_i(\rho u_j) \\
    \varepsilon\, n_{\ell\neq0}^1 = \frac{1}{3}\left[\frac{-(3\gamma+4\kappa)\left(Q^{ij}_\ell-\delta^{ij}/14\right)+2\gamma(p-1/2)\left(Q^{ij}_{\ell+1}-Q^{ij}_{\ell-1}\right)}{(3\gamma+4\kappa)^2+12\gamma^2\left(p-1/2\right)^2}-\frac{\delta^{ij}}{98\kappa}\right]\partial_i(\rho u_j),
\end{gather}
from which the viscosities \cref{eq:eta0,eq:etaH0,eq:zeta0} are computed.

\end{document}